\documentclass[a4paper]{article}

\usepackage[english]{babel}
\usepackage[utf8]{inputenc}
\usepackage{amsmath}
\usepackage{graphicx}
\usepackage[colorinlistoftodos]{todonotes}
\usepackage{cite}
\let\Item\item
\newcommand\SpecialItem{\renewcommand\item[1][]{\Item[\textbullet~\bfseries##1]}}
\renewcommand\enddescription{\endlist\global\let\item\Item}
\title{Quantum Clouds: A future perspective}

\author{Satish Bhambri}

\date{\today}

\begin{document}
\maketitle

\begin{abstract}
Quantum computing and cloud computing are two gaints for futuristic computing. Both technologies complement each other. Quantum clouds, therefore, is deploying the resources of quantum computation in a cloud environment to provide solution to the challenges and problems faced by present model of classical cloud computation. State of the art challenges faced by the cloud such as VM migration, data security, traffic management can be addressed by the quantum principles. But the merging of these two technologies have challenges of their own which need to be addressed before moving forward. What are those challenges and how does a quantum computer solve the cloud problems? The relation among quantum parallelism, superposition and flash crowd effect; Laundauer's principle and energy management; photon polarization principle and data security; these fascinating queries are addressed in the paper.
\end{abstract}

\section{Introduction}

It all began in 1981 with Richard P. Feynman quoting “ Can we simulate physics on a computer?” Well, though quite intriguing but answer to that question as of right now is obviously no and the only reason for that is a significant part of physics is Quantum physics which can’t be simulated on the computers present today, ie, the classical computers.\newline

Consider a system comprising of 30 particles. In Quantum physics we have variables in superposition equation to define a state of a particle. Consider that two variables are associated with each particle then, corresponding to a system of 30 particles we can have a state being described by 2 to the power of 30 variables, ie, a machine is in a state given by the combination of these variables. That amount of data can be stored with present classical computers. But what if number of particles go beyond 100. Correspondingly we’ll have 2 to the power of 100 variables. Now this can’t be simulated on a classical computer, since they don’t have this much memory and they never will !\newline

Feynman suggested “One can turn problem into an effect“. If Quantum physics is too rough for present computers may be we can use it to build better computers!\newline

Let's discuss a computing perspective.\newline 

One person who has been looking at the miniaturization of computers is Gordon Moore, also the co-founder of intel back in 1960’s. He predicted that number of components on a chip doubles every 18 months to two years\cite{Vitanyi}. Therefore, the smallest feature size on the Si chip has to decrease at the same rate. Hence, the Moore’s law. What’s amazing about Moore’s law is that you can predict what’s going to happen in time. Now, If we follow this law, less than 10 years from now, ie, as we approach 2020, the size of transistor will reduce to the size of an atom- the smallest component of nature. Thereby, here we enter the Quantum realm.\newline

Hence either way Quantum computing is inevitable future of computing.\newline

Another developing scenario of Computing is Cloud Computing, pointing towards a future in wherein we do not compute on local machines rather on centralized computing facilities driven by a third-party, controlling computing and storage facilities.\newline

Cloud computing is not a new term but has been into the market for a while now. Though being a strong futuristic Computing model it suffers from various research challenges\cite{zhang2010cloud} like :\newline

\SpecialItem
\begin{description}
  \item[] Automated Service Provisioning in clouds
  \item[] Migration of Virtual Machines
  \item[] Consolidating Servers
  \item[] Energy Management
  \item[] Traffic Management and Analysis
  \item[] Security of Data
\end{description}

 These are the issues preventing further progress in clouds. In this paper we explore a fascinating link between the two giants of futuristic computing "Cloud Computing" and the "Quantum Computing" and how they complement each other. A totally different way of addressing the issuesof clouds than the conventional methods. "Quantum Clouds" is thereby not just a solution to the challenges faced by cloud computing but a step forward in technical aspects of this computing model which quite unexpectedly marks the advent of quantum technology even more inevitable. What's even more fascinating is that quantum computing fills all the voids left by classical technology at every level of cloud architecture. For Instance a protocol which allows a machine's corresponding server execute the necessary quantum computation for it in a way that its inputs, outputs and computation remain perfectly private, eradicating its any requirements for private quantum computational power or memory, is given by the Blind quantum computing, which thereby complements the cloud security. The only requirement on the machine or client side is to prepare single qubits randomly chosen from a finite set and send them to the corresponding server with adequate quantum computational power.\cite{broadbent2009universal} \newline

\begin{figure}
\centering
\includegraphics[width=0.5\textwidth]{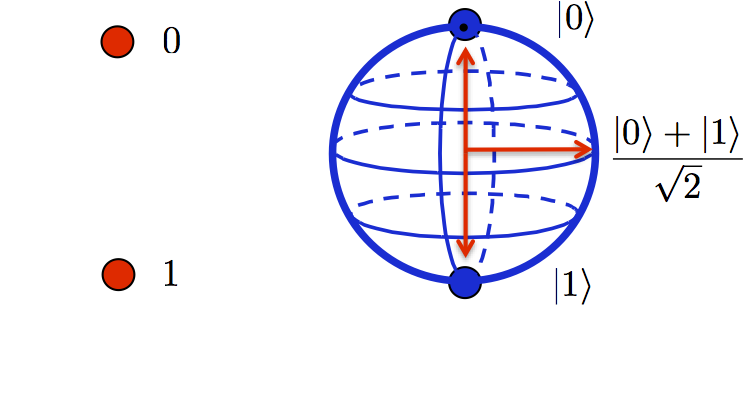}
\caption{\label{fig:image3}Qubit vs Classical bit\cite{stateofart}}
\end{figure}

With the advent of Quantum technology another issue that might need to be addressed is that due its high cost and maintainability, this quantum infrastructure will only be accessible to very few centres across the globe. The situation would be analogous to the one prevailing today with the limited resource provisioning of supercomputers to the selective clients. How will a client exploit this quantum resource without revealing the computation of the remote computer ? This Quantum computing challenge is fulfilled by the use of a cloud service wherein we use quantum infrastructure and deploy certain protocols of quantum computing. Hence, Cloud too complements the quantum computing.\newline

\section{Quantum Clouds}

Before Analyzing the indispensable role of a quantum computer in a cloud, we need to understand the basic architecture of cloud in order to determine where in that architecture does our quantum computer fits ?\newline

We have four layers in cloud which are as follows\cite{zhang2010cloud}:\newline
\SpecialItem
\begin{description}
  \item[] Application
  \item[] Platform
  \item[] Infrastructure
  \item[] hardware
\end{description}

Corresponding to these architectural layers we have a Service-driven business model, wherein these resources are provided as services on an on-demand basis.\cite{zhang2010cloud}\newline

We know that cloud offers services in three categories\cite{zhang2010cloud} :
\SpecialItem
\begin{description}
  \item[] Infrastructure as a Service
  \item[] Platform as a Service
  \item[] Software as a Service.
\end{description}
Corresponding to these services and resources, on an abstract level we have a following business model of cloud computing comprising of an End user, a Service Provider (SaaS), and an Infrastructure Provider(IaaS, PaaS).\newline
 
Now, in this model our Quantum Computer fits best on the most basic level, ie,  forming the core hardware of the cloud. Correspondingly our Quantum Algorithms and Automata
 form subsequent levels. Hence, the answer to our question.\newline

Now as per aim of this paper, we won’t be dwelling into the details of quantum computing or the cloud computing but only the relationships between these two that can help overcome the aforementioned limitations of the clouds. Also, we'll study the future perspective of the quantum clouds and the research challenges faced by this technology.\cite{singh2014quantum}

 \section{Automated service provisioning}
Starting with the Automated Service Provisioning, how would a quantum cloud be able to handle this research challenge of present clouds ? 
Automated Service Provisioning is about capability of cloud of acquiring and releasing resources on-demand. The objective of the a Service provider is to allocate and de-allocate resources from the cloud to satisfy its service level objectives (SLO’s). But the problem with this model is that to achieve high agility and to respond to rapid demand fluctuations in situations such as flash crowd effect, the resource provisioning decisions must be made online\cite{farkas2007automated}. This comes at the cost of increased management overhead, to dynamically reconfigure software components on instant demand. Users request resources like Virtual Machines from cloud infrastructure providers but time needed to configure them becomes a limiting factor, potentially offsetting the advantages of flexible computing infrastructure. Therefore, looking objectively, problem boils down to Space and time efficiency or, memory and speed of computation. The quantum computer increases the memory and speed exponentially, thereby, providing solution to any challenge that boils down to these two factors.

\section{Virtual Machine Migration and Traffic Management}

Let's take an example of a computing environment containing a server with a hypervisor and corresponding virtual machines, each running an OS and applications. The availability of the application would be significantly affected in the case if you need to bring down the server for maintenance, for instance, in order to increase the memory because in that scenario you'll have to shut down the software components and restart them after the maintenance. This VM migration in clouds is fascilitated by allowing you to move an entire virtual machine along with its corresponding operating system and applications, from one machine to another and continue operation of the virtual machine on the second machine. This advantage is unique to virtualized environments as you can take down physical servers for maintenance with minimal effect on 
running applications. This migration can be performed after suspending the virtual machine on the source machine, moving its attendant information to the target machine and starting it 
on the target machine. In order to minimize the downtime, the migration is performed while the VM is running (hence the name "live migration") and its operations are resumed on the 
target machine after all the state is migrated. 

Live migration, therefore, is done when the source machine goes down that can be due to either flash crowd effect or heating up of the components or various other technical gliches. The Quantum hardware enhances the capability of a cloud to handle any flash crowd at any time and to manage any traffic load effectively because of its immense processing and memory capabilities.A Quantum cloud can process multiple requests at a time owing to the principle of superposition of  states and Quantum parallelism. Quantum parallelism is phenomenon used to perform multiple computations at the same time. Quantum Interfernce thereby, afterwards is used to collect and combine the results. 
 
With the advent of cloud paradigm, live migration of multiple virtual machines have become much more pronounced and frequent. The live migration of
multiple virtual machines, unlike the single virtual machine migrations, faces many new challenges  such
as the migration conflicts due to the concurrent
migrations, migration failures due to the insufficient resources in
target machine, and the migration thrashing due to the dynamic
changes of virtual machine workloads. To maximize the migration efficiency in virtualized cloud data center environments, all the above issues need to be overcome.\cite{vmmigration2}
 
Live migration of a virtual machine can also be supported by the phenomenon of Quantum entanglement. Quantum entanglement can parallely store a dynamic state of a virtual machine on another qubit and we just need to apply a quantum OR gate so that in case of server failure of the original machine the control could fetch data from a parallel entangled qubit residing in another server. Hence, in this case a virtual live migration is taking place. This process also eliminates any discrepancies and failures associated with the classical live migration. 

Quantum paradigm also inculcates another phenomenon, Qauntum teleportation, by the virtue of which we can reproduce the same state of the qubits at a lab quite distant to the one we are operating in, avoiding any decoherence and without having any quantum channel in between, following a protocol known as quantum teleportation protocol. Therefore, this phenomenon cannot be simulated anyway in classical world, hence, such a kind of distant dynamic live migration of virtual machines can be accomplished in the quantum paradigm by deploying teleportation protocol.\cite{teleportation}

 \section{Server Consolidation and Energy management}

An effective approach to maximize resource utilization in a cloud environment, such that energy consumption is minimized is Server consolidation. Live Vm migration technology is used to consolidate virtual machines residing on multiple underutilised servers onto a single server, consequently remaining servers can be set to an energy-saving state.
The issue of optimally consolidating servers in a data center can be observed as analogous to the vector bin-packing problem, which is an np hard optimization problem.\cite{zhang2010cloud}

Shor's Quantum factoring algorithm solves the factoring problem which is also an np hard problem, thereby, laying up the groundwork for the solution of other np hard problems as well.\cite{shor}.A quantum hardware with 300 qubits is more powerful than all the supercomputers of the world put together. A silicon chip on a server consists of 3 million transistors on an average. The cost of running millions of transistors compared to few hundreds of qubits is very low once we develop appropriate technologies for a quantum computer. Therefore, server consolidation and energy management problems are addressed in quantum clouds.

Optimizing energy efficiency of servers is a considerable issue in cloud computing. Estimated data surveys have shown that the cost of energy management accounts for 53 percent of the total operational expenditure of data centres. Owing to the growing energy demands of the data centres, cloud infrastructure providers are under a significant amount of pressure to reduce energy consumption. 
For the same reason designing energy-efficient data centres has recieved considerable attention. This problem can be -approached from several directions in classical way in accordance with \cite{zhang2010cloud}.\newline 

\SpecialItem
\begin{description}
  \item[] Server consolidation
  \item[] Energy-aware job scheduling
 \end{description}

The key challenge in above directions is to achieve a good trade-off between energy savings and application performance. 

Quantum computing deploys reversible model model of computing Wherein the energy requirements are limited due to its reversible nature. \cite{limits}This was first proposed by Neumann and was later studied by Benett and Landauer.It has been shown that any computation could be carried out reversibly, thereby preventing any unnecessary dissipation of heat, which is the main cause of high energy requirements of cloud servers. Hence, this issue of energy management is aptly addressed in qauntum paradigm as it fundamentally exploits the reversible nature of computation.Refer Landauer's principle and reversible computation for further insight.\cite{landauer}

\section{Data security}
 
Data Security is among the most important research topic in cloud computing. Data security in quantum clouds is ensured using following protocols: \newline
SpecialItem
\begin{description}
  \item[]  Quantum Cryptography:\newline
 Quantum cryptography relies on the laws of quantum mechanics in order to ensure unconditional security. Contrary to quantum cryptography, conventional forms of cryptography either rely on a public key that everybody can access or on a private key.

The security of public key cryptography relies on the difficulty to realise an efficient algorithm to “crack” the communication. Being Unconditionally insecure due to the fact that no mathematical theorem forbids in these protocols, the client(sender) to build a clever revolutionary algorithm, or a quantum computer, that will allow him to crack such codes.
Quantum key dirstribution technique, developed by Charles Bennett and Gilles Brassard, in the 1980s, deploys a protocol using quantum principles in order to securely distribute random bits between the sender and the reciever of a two way communication systems. 
BB84, as the quantum cryptography protocol by Bennet and Brassard is known as, is absolutely secure against discovery during transmission against any classical methods.\cite{qkd}
Quantum principles deployed in BB84 protocol to achieve security against eavesdropping are the photon 
polarization principle and Heisenberg’s uncertainty 
principle. 

 RSA public key cryptosystem, whose security relies on the hardness of factoring—a problem that a quantum computer can solve efficiently. Shor's factoring algorithm, a quantum algorithm, forms an integral part of this quantum phenomenon. The time a classical circuit takes to factor a number with n digits grows exponentially with n, implying that for numbers with fairly large number of digits a classical computer takes very long time, thereby, limiting their practical feasibility in RSA cryptosystems, which take advantage of this difficulty and consequently, a large amount of information is protected by large semi prime numbers.(A semi prime number refers to product of two primes)

Private key cryptography can be unconditionally secure if encryption techniques such as the ‘one time pad’ are deployed but the weakness of these techniques is the secure transmission of the key from Alice(sender) to Bob(reciever), whilst at the same time they are using cryptography because  classical transmission channels cannot be relied upon.

The unconditionally secure transmission of a random binary key between Alice and Bob, is enabled by the quantum cryptography, in order to solve the aforementioned dilemma, and hence is referenced as Quantum Key Distribution (QKD).\cite{qkd} Basically, the transmission security is ensured by the application of no-cloning theorem which explicitly forbids the exact reproduction, or cloning, of a quantum system without disturbing it, thereby enabling Alice and Bob to detect the presence of a potential eavesdropper.

For further insight in quantum cryptography, refer \cite{cryptography}

	\item[] Blind Quantum Computation:\newline
Quantum infrastruture preserves the privacy of a computation along with the exponential speedup in computation. Blind quantum computing is the one wherein  the input, computation, and output all remain unknown to the computer. The blind computation deploys various quantum transformation including one and two qubit gates and the Deutsch and Grover quantum algorithms. As discussed further in detail in \cite{blind1}, blind quantum algorithm exploits th conceptual framework of measurement-based quantum computation which enables a client to delegate a computation to a quantum server. The client needs to be able to prepare and transmit individual photonic qubits. This is quite rudimentary for unconditionally secure quantum cloud computing and might become an indispensable ingredient for real-life applications, especially when considering the challenges of making powerful quantum clouds.\cite{blind1}

Universal blind quantum computation:
While dealing with quantum clouds, one aspect is quite commendable that computation power of each client can be raised exponentially without actually carrying out or without having any such quantum computational poweror memory on the client side. The Universal blind quantum computation adds to this aspect by allowing a client to have a server carry out quantum computation for him given the fact that his inputs, outputs and computation remain perfectly private. The client machine needs to be only capable of preparing qubits chosen from a given finite set and pass them on to the server which has sufficient resources following which two way classical communication occurs enabling client to give single qubit measurement instructions to the server which further depend on the outcomes of previous measurements.
 
The aforementioned protocol works for inputs and outputs that are either classical or quantum leading to an authentication protocol allowing the client to detect an interfering server(eavesdropper). The result is generalised to the setting of a purely classical client who communicates classically with two non-communicating entangled servers, in order to perform a blind quantum computation.Therefore, we have a quantum protocol, which is the universal fault tolerant scheme detecting a cheating server, as well as does not requiring any quantum computation whatsoever on the client's side. \cite{blind}

  \item[] Quantum Data hiding

 Quantum data hiding protocol in quantum clouds is another feature that strengthens the data security by hiding classical data among parties who are restricted to performing only local quantum operations and classical communication (LOCC). Quantum data hiding process hides one bit between two parties using Bell states(state of a system of two qubits, where the composite state is an equal superposition of 00 and 11 or considering spins os the electrons in a two electron system wherein electrons represent the qubits), and derives upper and lower bounds on the secrecy of the hiding scheme. Remote preparation of the quantum states in the communication between client and server ensures enhanced data security. Discussion of that protocol is beyond the scope of this paper. \cite{remote}
 
 This technique provides an explicit bound which shows that multiple bits can be hidden bitwise and therefore, gives a preparation of the hiding states as an efficient quantum computation that uses at most one qubit of entanglement. This scheme for quantum data hiding can be used in a conditionally secure quantum cloud architecture. To dig further in, refer \cite{data}
 
Hence, Its quite evident that dawn of the quantum realm in cloud technology will inevitably, overcome the present data security challenge.
Despite innumerable benefits of quantum clouds, this paradigm also suffers from some quite significant challenges.

\section{Limitations of Quantum Clouds}
Quantum clouds inherit their limitations from the basic principles of Quantum mechanics.Some limitations are intrinsic to the nature of Quantum paradigm itself, for instance, No cloning theorem, heisenberg's uncertainity principle and others are practical or constraints of our present technology such as decoherence. Other limitations are induced by some present quantum algorithms, for instance, not all quantum algorithms are exponentially superior to classical ones as expected but some like searching the database with Grover's algortihm can provide quadratic speedup over the classical at the most\cite{Vitanyi}\cite{aronson}.

\begin{description}
  \item[] Quantum decoherence\newline
  If the quantum processor is not perfectly isolated from the environment, it couples with the latter one and the quantum states are modified or stating it otherwise, original superposition of states is adulterated leading to the phenonmenon known as quantum decoherence.\cite{Gruska}
  We avoid the intrinsic decoherence properties of the circuit by getting rid of the junk bits genereated in quantum circuits by deploying a similar reversible circuit and copying the required bits to workspace using the CNOT gate. This forms the basic step of some very prominent quantum algorithms, for instance, Simon's algorithm, Shor's factoring algorithm and Grover's algorithm.
  
 Therefore, Quantum dynamics of the environment surrounding the processor are quite significant and relevant to quantum cloud as they tend to entangle the qubits of the system to the environmental quantum states thereby irreversibly destroying the information in the superposition of quntum processor states hence, making decoherence as one of the fundamental limitation kin the quantum realm\cite{Gruska}.
 \item[]Heisenberg's uncertainity principle\newline
 It states that accurate measurement of one observable makes the value of another observable less certain. In terms of qubits this principle can be stated as the product of spread of a qubit in sign basis and in standard basis cannot be less then square root of 2.This implies that we cannot determine the sign value and the bit value of a qubit accurately simultaneously.
 
 \item[]No cloning theorem\newline
 It states that an unknown quantum state cannot be cloned.Owing to this theorem, quantum error correcting codes seem infeasible, though it also presents an advantage that due to the improbable nature of copying of quantum information, we have quantum key generation systems which are unconditionally secure. Its proof is beyond the scope of this paper. Refer \cite{Gruska} for further insight. The use of this limitation is indispensable in quantum clouds as can be deferred from \cite{nocloning}.
 \item[] Holevo theorem\newline
 It limits the amount of quantum information that we can extract from a quantum state due to the measurement principle of the quantum mechanics. We are only able to access that state of the superposition of the corresponding qubits which result as the outcome of measurement and rest all are destroyed. According to Holevo theorem, corresponding to n qubit state register we can encode and decode safely only n bits though using dense coding we can encode n bits in m(less than n) qubits\cite{Gruska}.

 To deal with such limitations, Quantum fault tolerance(topological or brute force) and Quantum error correction methods are deployed and these are still under research arena.For further insight refer\cite{tolerance2} \cite{tolerance3}\cite{tolerance1}

\end{description}

\section{Research challenges in Quantum Clouds}
Quantum technology is in yet its poineering stages, therefore having a significant amount of challenges which need to be addressed before it can be deployed on a larger scale.\cite{Gruska}

\begin{description}

  \item[] Quantum Measurement\newline
  The measurement dimension of quantum realm reveals only a particular part of the state which is a complete superposition of the probable states of the corresponding qubits involved with normalized amplitudes with various probabilities. Thus exact state of a system cannot be determined. The evaluation of the cost of quantum computations in accordance with the quantum complexity theory is hindered by this limitation \cite{Gruska}. Its significance can increase as the number of required measuring devices increase for some polynomial time quantum algorithms.
  Despite it being a research challenge, it can be used to battle another quantum cloud limitation of decoherence. Entanglement protection scheme makes use of the quantum measurement and can effectively circumvent even entanglement sudden death (ESD)\cite{esd}.

  \item[] Multipartite Entanglement\newline
  Understanding and manipulation of multipartite entangled states is indispensable for distributed quantum computing and networks and therefore, plays an important role in quantum error correcting codes\cite{Gruska}. Since, entanglement cannot be shared among different particles in accordance with \cite{Gruska}, therefore its a flamboyant research challenge to ascertain the limitaions and possibilities of multipartite entanglement.

  \item[] Quantum Information Processing \newline
  The challenge in this domain is to discover what is actually beyond "shor's and Grover's". How to find quantum algorithms for the problems like graph isomorphism, approximate shortest vectoras well as nonabelian hidden subgroup problem. The challenge is to address a rudimentary question of whether "fundamentally new" quantum algorithms remain to be discovered \cite{scott}.
  In the area of computation theory, the challenge remains to determine the computation power of qauntum cellular automata as compared to quantum turing machine.\cite{Gruska} Quantum clouds need to address these challenges for the efficient implications of a quantum infrastructure in cloud environment.
  
  \item[] Quantum cloud processors\newline
  Processors that could carry out quantum computation on a wider and reliable scale have been under the scanner of immense technological research \cite{processor}. We have experimentally demonstrated the entanglement,teleportation and implementation of quantum algorithms on 2 to 7 qubit processors amon which only superposition have been demonstrated on 7 qubit processors.Various techniques proposed and used in developing these qubit processors are trapped ions, liquid state NMR, electron spin transistors, but the adversities associated which challenge further development are setting up of an initial state, stabilizing quantum memory,performing measurement readouts, battling decoherence. The scope of qubit processors needs to be enlarged for its practical implementation in Quantum clouds so that they can shoulder the burden of the information to be processed. \cite{processor}\cite{Gruska}\cite{scott}.

 \item[] Quantum-classical Interface protocols\newline
  This challenge is yet a novice considering the fact that once we have implemented a basic quantum computer and devised the requisite quantum algorithms, next step would be to create an interface between classical and quantum infrastructure as with quantum clouds, the client will be having classical infrastructure.Networking protocols that could incorporate quantum levels
along with classical levels along with the implementation of concepts of blind quantum computing and quantum cryptography need to be developed.

  There are many other challenges and limitations to Quantum paradigm, some of which can be explored in \cite{scott}.

\end{description}

\section{Epilogue}
 
 Despite some research challenges, which yet need to be addressed, quantum clouds inevitably mark the dawn a computing era with unimaginable possibilities and computing ways which might as of right now seem grandiose. As feynman said we have turned the problem of quantum principles into an effect to incorporate those principles into computing realm.The success and excitement of this forseeable futuristic technology is even enhanced by the evidence that it is filling up the voids left by classical computing models.
 
 Cloud computing is among the frontier models of classical computing and deploying quantum paradigm in this arena is of utmost importance.The state of the art problems and challenges currently faced by the clouds, for instance, automatic service provisioning, VM migration, traffic management,Server consolidation and data security  are addressed by the introduction of quantum mechanical principles.
  
 In layman terms, whenever we think of future technology, we manifest a picture wherein we have thin and extremely revolutionary computing gadgets with enormous computation power. Quantum clouds is basically a step in that direction which propagates the jist of computing in which exponential computation power will reside at the server(a quantum cloud model)
 and device will just be requiring a connection to that.
 
 Though having great advantages, this quantum clouds too suffer from "drawbacks" as of now. The actual implementation is still way down the road.We need to address the problems of decoherence, measurement, development of quantum processors, and discover effective ways for information processing.
 
 Ultimately, research is still going on at an appreciable pace, hence the question now is not IF quantum clouds will be there, but the actual question is WHEN ?
 
 \section{Acknowledgements}
 I would like to thank
 	Mr Umesh Vazirani for inspiring me and motivating 	me in this field. It was after watching his lectures 	online, i developed passion in this arena. \newline
 Ms.Karamjeet Kaur Cheema, for her guidance in 		cloud computing.\newline
    Mr Ajay Kumar Laura, for his guidance in theory of 		computation.\newline
    Mr Anil Kumar Verma , for his support and driving 	me towards research.

\end{description}
\bibliographystyle{plain}
\bibliography{q1,q2,q3,q4,q5,q6,q7,q8,q9,q10,q11,q12,q13,q14,q15,q16,q17,q18,q20,q21,q22,q23,q24,q25,q26,q27}

\end{document}